\documentstyle[aps,floats,epsf,twocolumn]{revtex}

\begin{document}

\title{Theory of spin response in underdoped cuprates as
 strongly fluctuating d-wave superconductors}

\author{Igor F. Herbut and Dominic J. Lee}

\address{Department of Physics, Simon Fraser University, 
Burnaby, British Columbia, Canada V5A 1S6} \maketitle

\begin{abstract}
We study the spin dynamics in underdoped cuprates at low temperatures 
by considering them as quasi two dimensional
d-wave superconductors with strong quantum
phase fluctuations. An effective theory of spin degrees of freedom
of nodal quasiparticles coupled to vortex defects in the phase of the
superconducting order parameter is formulated. It represents the minimal
extension of the $QED_3$ theory of the pseudogap phase 
into the superconducting region. The theory predicts 
a single superconductor-spin density wave (SC-SDW) phase transition,
without coexistence between the two phases. At the transition,
which may be fluctuation induced first-order, 
vortices condense (and SC is lost) and the chiral symmetry for fermions
simultaneously breaks (and SDW is formed).
We compute the spin-spin correlation function in the 
fluctuating superconducting state, and explain the 
evolution of the spin response with energy in underdoped
$Y Ba _2 Cu _3 O_{6+x}$ observed in neutron scattering experiments.
In particular, we predict that at very low frequencies $\omega\sim
(1/10) \omega_{res} $, with $\omega_{res}$ being the energy of the
`resonance' at $\vec{Q}= (\pi,\pi)$,
(weak) spin response should become narrowly peaked at four diagonally
incommensurate wave vectors that span between the nodes of the
superconducting order parameter.  These peaks represent the inherent 
collective mode of the phase fluctuating d-wave superconductor, the condensation
of which would bring about the SDW order in the pseudogap phase.
Our interpretation of the resonance suggests that it should become more
elusive in the superconductors with lower $T_c$, in accord with its'
conspicuous absence in $La _{2-x} Sr _x Cu O_4 $.
\end{abstract}

\vspace{10pt}

\section{Introduction}

Underdoped high temperature superconductors are highly anisotropic,
quasi two-dimensional materials, in which thermal and quantum
fluctuations of the phase of the superconducting order parameter (OP)
should be important well below the pseudogap temperature.
It has recently been shown that a  
phase fluctuating d-wave superconductor (dSC) at zero temperature  ($T=0$)
is inherently unstable towards the formation of the very weak 
spin density wave (SDW) state with loss of phase coherence \cite{herbut},
\cite{babak}. This follows from the observation that the standard
Fermi liquid theory for the BCS d-wave quasiparticles,
besides the usual spatial
symmetries, also acquires an additional continuous (`chiral' \cite{herbut})
symmetry at low energies. This symmetry
becomes dynamically broken when the superconducting phase
coherence is lost via proliferation of vortex defects.
The theory of the phase incoherent (pseudogap) state in this
formulation becomes closely related to
the three dimensional quantum electrodynamics ($QED_3$) \cite{franz},
in which the  coupling constant (`charge') is proportional to the vortex
condensate.
In the crudest approximation, the fluctuations of the gauge-field
may then be neglected in the superconducting state, which leaves
quasiparticles as sharp low energy excitations. When the vortices
condense and the charge in the $QED_3$ is finite, the
chiral symmetry of the dSC becomes dynamically broken, in
analogy with the well known phenomenon in particle physics \cite{pisarski}.
In the present context this translates into weak SDW ordering,
with confined spin-1/2 (spinon) excitations \cite{herbut}.

Several conceptual and practical issues of direct relevance to cuprates arise 
naturally in such an approach to underdoped high temperature superconductors.
Can the SC and the SDW long-range orders coexist? Such a coexistence
appears rather generically in various mean-field treatments
of microscopic models of cuprates \cite{tremblay},
\cite{brinckmann}. Several intriguing
 recent experiments \cite{sonier}, \cite{mook}, \cite{sidis},
 also suggest the coexistence of the two inimical orderings. If there is
 a single dSC-SDW quantum phase transition, on the other hand,
 what should be its universality class? What are the effects of
 the virtual vortex fluctuations {\it inside} the superconducting state?
 These questions all require a better understanding of the
 quasiparticle-vortex interaction, particularly in the superconducting state.

 A theory of low-energy d-wave quasiparticles coupled
 to phase fluctuations of the superconducting OP which puts
 the earlier physical ideas of \cite{herbut}, \cite{franz} into a
 consistent field-theoretical framework has recently been proposed
 in \cite{dom}, \cite{igornew}. One result of that study 
 is that, at least within a certain approximation \cite{igornew},
  in the superconducting state spin of quasiparticles may be considered
 asymptotically decoupled from their charge at low energies. Physically
 this is just what is expected, and may be understood as being a consequence
 of the long range phase coherence in a singlet superconductor \cite{kivelson},
 \cite{balents}: the 
 condensate screens perfectly the charge of quasiparticles, effectively
 transforming them into neutral, spin-1/2 excitations.

 In this paper we wish to take advantage of this state of affairs and 
 consider a simpler effective $T=0$ theory (Eq. (1)) for the 
 spin sector of the fluctuating dSC only. We assume that
 in underdoped high temperature superconductors the amplitude
 of the superconducting OP may be considered frozen well below
 the pseudogap temperature $T^*$, but that the quantum fluctuations
 of its phase will inevitably arise sufficiently
 near half filling when the interaction between electrons
 becomes strong. In the phase incoherent (pseudogap)
 state our effective theory reduces to the previously studied
 $QED_3$ \cite{herbut}, \cite{franz}, with its concomitant SDW
 instability. Deep in the  dSC, 
 on the other hand, the theory becomes equivalent to the three
 dimensional Thirring model, with an (irrelevant) short range interaction
 between the electrically
 neutral spin-1/2 excitations, which we will, as traditionally,
 call spinons. The effective theory in Eq. (1) connects, and describes 
 the region in between, these two limits, and establishes a physical link 
 between the phase fluctuations and the
 magnetic excitations in underdoped cuprates.

 Having the low energy theory for the spin degrees of freedom
 we proceed to derive several interesting consequences from
 it. First, we show that the
 condensation of vortices (i. e. the loss of superconductivity) and
 the  breaking of the chiral symmetry for fermions (the
 SDW instability) in the theory coincide. The region of
 coexistence between the SC and the SDW long-range orders is not found.
 Second, we find that the quantum dSC-SDW phase transition could be
 fluctuation-induced first-order, in agreement with our earlier study
 of the full theory that included charge \cite{dom}. Finally, and maybe most
 importantly, we calculate the spin dynamics {\it induced} by vortex fluctuations 
 inside the superconducting state, in the leading approximation.
 The detailed evolution of the spin response with energy observed in the
 neutron scattering experiments on $YBa_2 Cu_3 O_{6+x}$  (YBCO) 
 \cite{dai},\cite{fong}, \cite{arai} is explained in terms of four
 diagonally
 incommensurate `mother' peaks, that should become discernible at lowest 
 frequencies. These weak and narrow peaks are centered on
 the wave vectors that span between the diagonally opposite
 nodes of the superconducting OP, and represent the collective particle-hole
 mode of the phase fluctuating dSC.
 The SDW instability obtained previously with the $QED_3$ theory can be
 understood qualitatively as the condensation of this collective mode. 
 The famous `resonance' at the commensurate wave vector $\vec{Q}=(\pi,\pi)$
 is here obtained as an overlap between the all four mother peaks, which
 dominates in the spin response within certain window of energies. This
 energy window shrinks with the superconducting transition
 temperature $T_c$, which may explain why the resonance has not been
 observed in $La_{2-x}Sr_x Cu O_4 $ (LSCO) \cite{fujita}.

   The paper is organized as follows. In the following
    section we introduce the
   effective theory of spin of quasiparticles, and discuss its physical
   motivation and the formal justification. In sec. III we show how
 our theory avoids the  coexistence of the SC and the 
   SDW orders. In sec. IV we briefly discuss the nature of the
   dSC-SDW quantum phase transition. Spin response is discussed 
   in sec. V. The discussion of
   our results and the relation to  other approaches
   is presented  in the concluding section. 

\section{Theory of spinons and vortices}

Consider the quantum mechanical ($T=0$) action for the low energy
quasiparticles in the two-dimensional phase fluctuating dSC,
$S= \int d^3 x {\cal L}$, $x = (\tau, \vec{r})$, $\tau$ is the imaginary
time, and
\begin{eqnarray}
{\cal L} = \sum_{i=1}^N
\bar{\Psi}_i \gamma_\mu (\partial_\mu - i a_\mu) \Psi_i
+ \frac{i}{\pi} \vec{a} \cdot (\nabla \times \vec{A}) + \\ \nonumber
| (\nabla - i \vec{A} )\Phi|^2 + \mu ^2 |\Phi|^2 + \frac{b}{2} |\Phi|^4.
\end{eqnarray}
The fluctuating complex field $\Phi$ describes the vortex loops,
and $\langle\Phi \rangle = 0$ implies that all such loops are of
finite size, and thus the superconducting phase coherence.
$\Phi$ should be understood as being {\it dual} to the
standard superconducting OP, and $\langle\Phi \rangle \neq 0$
means that vortex loops proliferated and that superconductivity is lost
\cite{igorjpa}.
Two ($N=2$) four-component Dirac fermions describe
the gapless, neutral, spin-1/2 (spinon) excitations near the four nodes
of the superconducting order parameter at $\pm \vec{K}_{1,2}$,
one Dirac field for each pair of
diagonally opposed nodes, as defined previously in \cite{herbut}.
$\vec{a}$ is the  gauge-field that results from 
absorbing the singular part of the superconducting phase due to vortices
into the spinon fields \cite{franz}, \cite{herbut}.
$\gamma_\mu$, $\mu =0,1,2$ are the Dirac gamma matrices
\cite{herbut}, and $\{ \gamma_\mu,\gamma_\nu  \} = 2\delta_{\mu \nu}$.
Finally, $\vec{A}$ is an 
auxiliary (Chern-Simons) field which facilitates the statistical
spinon-vortex coupling, as will be explained shortly.  
Tuning parameter $\mu^2$ controls the magnitude of vortex fluctuations, 
and we assume it to be related to doping, $\mu^2 \propto (x-x_c)$, with
$x_c$ being the critical doping in the underdoped regime.
The coupling $b>0$ describes the short-range repulsion between the vortex
loops. We work in units in which $\hbar=k_B=1$,
and we have set  the two characteristic velocities of the
nodal quasiparticles $v_F=v_\Delta=1$ in (1), for simplicity.
The amplitude of the superconducting order parameter is assumed frozen,
and Eq. (1) is supposed to describe only the spin and the
vortex degrees of freedom well below the pseudogap temperature $T^*$,
as in the previous work \cite{herbut}.

One may physically understand the form of the theory (1) as follows
\cite{lannert}: the exact integration over the gauge-field $\vec{a}$,
which appears only linearly in (1),  leaves the spinon-vortex coupling
$i \vec{A} \cdot \vec{J}_{\Phi}$, where $\nabla\times \vec{A}= \pi 
\vec{J}_{\Psi}$, and $\vec{J}_{\Psi}$  and
$\vec{J}_{\Phi}$ are the spin and vortex current densities,
respectively. Vortices and spinons therefore see each other as
sources of a fictitious magnetic flux, and circling around a vortex with a spinon
(or vice versa) leads to a phase change of $\pi$. This is precisely
the {\it statistical} (Aharonov-Bohm) part of the interaction
between vortices and quasiparticles \cite{franz}, \cite{herbut},
which couples only to spin.
The regular part of the phase, which provides the standard Doppler
shift of quasiparticle energies and couples only to charge,
can be `gauged away' \cite{balents} at the price of
introducing an additional, ultimately {\it irrelevant} interaction term.
Absorbing this regular part of the phase into quasiparticles converts them
into spinons \cite{balents}, represented by $\Psi$ in Eq. (1)
\cite{herbut}. The final
point is that at low energies and in the superconducting state the
spin of quasiparticles may be considered decoupled from its charge
\cite{igornew}. Essentially, this is because one can consider
the superconducting condensate to carry all the charge in the
phase coherent state.
Similar conclusion  may also be reached within the gauge theory
formulation of the t-J model \cite{dhlee}.

The separation between spin and charge allows one to study spin excitations
separately from the charge in the fluctuating dSC at low energies.
Eq. (1) may therefore be
understood as an {\it effective} theory for the spin of quasiparticles.
One can also derive it from the full theory which includes
charge \cite{dom}, \cite{igornew} by simply omitting
the additional Chern-Simons field for the charge degrees of freedom.

Finally, in writing (1) the anisotropy between the Fermi  and 
the OP related velocities, $v_F \gg v_\Delta$, 
and all the local interactions between quasiparticles \cite{babak}
have been neglected, as irrelevant at low
energies \cite{lee}, \cite{vafek}. We will, however, find it
necesary to restore the velocity
anisotropy later when we discuss the spin response, 
where this feature will turn out to play an important role.

Although in this work we are concerned with the vortex fluctuations
in the superconducting state, which is a complementary problem to the
one studied in the 
previous papers on the $QED_3$ theory of
the pseudogap phase \cite{herbut}, \cite{franz},
let us nevertheless mention how the theory (1) is related to the $QED_3$
in the mean-field approximation for the vortex
field. Entirely neglecting the fluctuations in $\Phi$, in the superconducting
phase ($\langle \Phi \rangle = 0 $), the integration over $\vec{A}$
would constrain $\nabla \times \vec{a} =0$, and spinons become free.
When vortices condense ($\langle \Phi \rangle \neq 0$), on the other
hand, the gauge-field $\vec{A}$ acquires a mass via Higgs mechanism,
$|\langle \Phi \rangle| ^2 \vec{A}^2 $. The Gaussian integration over
$\vec{A}$ produces then the Maxwell term for $\vec{a}$,
$\sim (\nabla \times \vec{a})^2 / |\langle \Phi\rangle| ^2$. Together 
with the Dirac Lagrangian this constitutes the 
$QED_3$ for spinons, in which the fermions' chiral symmetry generated
by $\gamma_3$ and
$\gamma_5$ is dynamically broken by the generation of the small mass term 
 $\sim M \bar{\Psi}_i \Psi_i$ \cite{pisarski}. In the present context such a
mass $M\sim \langle {\bar \Psi}_i \Psi_i\rangle \sim |\langle \Phi \rangle|^2 $
is proportional to the weak SDW OP, with the (incommensurate) ordering
wave vectors $\pm 2\vec{K}_{1,2}$ that connect the diagonally opposed nodes
of the superconducting OP \cite{herbut}.  

 Our aim will therefore be to understand how the theory (1) interpolates
 between the limits of $QED_3$ in the pseudogap, and the non-interacting
 theory of spinons in the superconducting phases. In particular, we
 want to understand the evolution of spin dynamics with progressive
 quantum disordering of the dSC, which is assumed to correspond to
 underdoping. We begin with the issue of possible
 coexistence of the SDW and the dSC orderings. 

\begin{figure}
\centerline{\epsfxsize=7cm \epsfbox{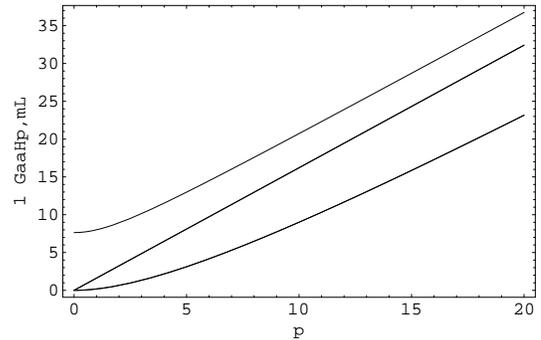}}
\vspace*{1em}

\noindent
\caption{Schematic behavior of the inverse gauge-field propagator
$1/G_{aa}^0(p,m)$ as a function
of the momentum in the SC ($m^2 >0$), critical ($m=0$),
and the pseudogap ($m^2 <0$) phases
of the theory (1), from top to bottom. At large momenta the behavior is always linear.
At low momentum, however, the statistical gauge
field is massive ($\sim m$) in the SC phase, and has a Maxwell term
($\sim p^2$ ) in the pseudogap (incoherent) phase.} 
\label{f1}
\end{figure}

\section{Coincidence of the SC and the SDW transitions}

In the mean-field approximation, the vortex condensation and the
SDW transition in the theory (1) coincide.
It is not obvious that this feature survives the
inclusion of the fluctuations, and it is conceivable
that the SDW transition  may occur within the
superconducting phase \cite{herbut}, \cite{tremblay},
\cite{brinckmann}, \cite{pereg}. To examine this issue we
first ask what would be the {\it exact} propagator for the gauge field
$\vec{a}$ if there were no fermions, i. e. when $N=0$ in (1).
We then approximate the interacting  
action for $\vec{a}$ that would result from the integrations over $\Phi$ and
$\vec{A}$ with the effective Gaussian term that reproduces that exact propagator.
In the Landau gauge such a propagator is 
\begin{equation}
G^0 _{aa,\mu \nu} (p,m) = \frac{\pi^2 \Pi_{AA} (p,m)}{p^2}
(\delta_{\mu \nu}-\hat{p}_\mu \hat{p}_\nu ),
\end{equation}
where $\Pi_{AA} (p,m) (\delta_{\mu \nu}-\hat{p}_\mu \hat{p}_\nu )$
is the transverse current-current correlation function in the
$|\Phi|^4$ theory. Here $m^2 =\mu^2 +O(b)$ is
the fully renormalized "mass" of
the vortex field. In general, $\Pi_{AA}(p,m) = n p F_ + (m/p)$,
for $m^2 >0$ (superconductor). To the lowest order in $b$
\begin{equation}
F_+ (z) = \frac{1}{8\pi} [ (4z ^2 +1) \arctan(\frac{1}{2z}) - 2z] +O(b).
\end{equation}
Here  we generalized our model to the one with $n$ complex
vortex fields, $n=1$ being the case of physical interest.

 We are now in position to
  study the chiral symmetry breaking for fermions with (2)
 serving as the bare (without fermion polarization) gauge-field
 propagator in the
 $QED_3$. Consider the standard large-N Dyson equation \cite{pisarski}
 for the fermion self-energy
 \begin{equation}
 \Sigma(q) = \frac{1}{4} Tr \int \frac{d^3 p}{(2\pi)^3}\gamma_\mu
 \frac{ G_{aa, \mu \nu} (p-q ,m) \Sigma(p)}{ p^2 + \Sigma(p)^2 }
 \gamma_\nu
 \end{equation}
 Here $[G _{aa,\mu \nu} (p,m)]^{-1} = \Pi^F _{aa,\mu \nu}(p) +
 [G^{0} _{aa,\mu \nu} (p,m)]^{-1}$,
 with the one-loop fermion polarization $ \Pi^F _{aa,\mu \nu}(p)= 
 (Np/8)(\delta_{\mu \nu} - \hat{p}_\mu \hat{p}_\nu)$.
 It is useful to consider first the point of the 
 superconducting phase transition $m=0$, where Eq. (3) implies 
 $\Pi_{AA} (p,0) =  np(1 /16 +O(b_c ) )$, where $b_c$ is the
 fixed point value of $b$ in $d=3$ \cite{cha}. Linear dependence
 of $\Pi_{AA}(p,0)$ on momentum is an exact result
 \cite{herb-tes}. Inserting this into the Dyson equation  we find
 that at $m=0$ the effect of vortices is only to increase  
 the coefficient $N\rightarrow N_{eff}=N+ 128/(\pi^2 n) + O(b_c)$.
 Recalling that the non-trivial solution of the Dyson equation (4)
 exists only for $ N_{eff} < N_c = 32 / \pi^2 $ \cite{pisarski},
 for $N=2$ we find that chiral symmetry
 at $m=0$ would be already broken only for
  $n > n_c = 10.44( 1+ 32/(9\pi^2 n ))$,
  where we have also included the known
  $O(b_c)$ correction to $\Pi_{AA}(p,0)$
 in the large-$n$ approximation \cite{cha}.
 Since $n=1$ in the physical case we conclude that right at the
 superconducting critical point SDW order is most likely absent.

 Little further thought shows that, if correct,
 the above result implies that the SDW order
 is absent for all $m^2 >0$. Indeed, this is to be expected, since
 $G^0 _{aa} (p,m)< G^0 _{aa} (p,0)$, and
 $\vec{a}$ is only {\it stiffer}  in the superconducting
 phase. To prove this, assume that for $m^2 >0$ a
 non-trivial solution of the Eq. (4), $\tilde{\Sigma}(q)$, does exist.
 Such a solution would then satisfy
 \begin{equation}
 \tilde{\Sigma}(q) <  \frac{1}{4} Tr \int \frac{d^3 p}{(2\pi)^2}\gamma_\mu
 \frac{ G_{aa,\mu \nu} (p-q,0) \tilde{\Sigma}(p)}
 { p^2 + \tilde{\Sigma} (p)^2} \gamma_\nu.
 \end{equation}
 On the other hand, we already established that there is only a trivial
 solution of the Dyson equation at $m=0$. This means that assuming a
 candidate function $\Sigma(p)$ and inserting it under the integral
 in the Dyson
 equation (4) for $m=0$ will produce only a {\it smaller} new $\Sigma(q)$,
 since under iterations the physically acceptable self-energy must approach
 the trivial solution. This  further implies that any
 $\Sigma(q)$ from the domain of attraction of the trivial solution
 at $m=0$ has to satisfy the inequality opposite to (5). 
 A non-trivial $\tilde{\Sigma}(q)$ at $m^2 >0$ therefore can not
 exist if it did not exist at $m^2 =0$.

   For $m^2 <0$ superconductivity is lost, and $\Pi_{AA} (p,m)
 = n p F_- ( |\langle \Phi \rangle|^2 /p )$, with $F_-(z)
 \rightarrow 1/16$ for $z \ll 1$, and $F_- (z)=2 z$, for $z \gg 1$
(Higgs mechanism), to the leading order.  The behavior of the gauge
field propagator is depicted in Fig. 1. 
Since $G^0 _{ aa} (p,m)> G^0 _{ aa} (p,0)$
 for $m^2 <0$, the above argument no longer applies. In fact, it is easy to
 see that there is {\it immediately}
  a non-trivial solution of the Eq. (4) when
 $m^2 < 0$. Since $\Sigma(q)>0$ only for  $ q \sim |\langle \Phi \rangle|^2$
 and smaller, $|\langle \Phi \rangle|^2 $ serves as the effective ultraviolet
 cutoff in the Eq. (4). The Dyson equation then reduces to the
 standard one in the $QED_3$
 \cite{pisarski}. For $N=2 < N_c $,  $\Sigma(q=0) \sim
 |\langle \Phi \rangle|^2$, and the chiral symmetry is dynamically broken.
 There is no intermediate (quantum disordered)
 phase in between the SDW and the dSC in the theory (1), unless the exact
 value of $N_c $ in the $QED_3$ is actually less than two \cite{cohen},
 \cite{hands}.

\section{Nature of the transition}

   The above argument strongly
   suggests that there is a single dSC-SDW transition
 in the theory (1), but does not say anything about its nature. This
 issue in the full theory with both spin and charge was addressed in
 \cite{dom}, and here we will provide a brief alternative calculation in 
 our effective theory that supports the earlier results. To
 study the phase transition
  it is better to proceed in the opposite direction,
 and integrate over the fermions first. This gives dynamics to $\vec{a}$
 via the fermion polarization bubble.
 If we neglect the quartic and the higher
 order terms and perform the Gaussian integral over $\vec{a}$, the result is
 the Maxwell-like term for $\vec{A}$:
 $\langle A_\mu (p)A_\nu(-p) \rangle = (N\pi^2 /8|p|)
 (\delta_{\mu\nu}- \hat{p}_\mu \hat{p}_\nu) $.
 Note that since the inverse of the
 above average is non-analytic at $p=0$,
 it can not renormalize \cite{herbut1}, and therefore the number of fermions
 $N$ represents an exactly {\it marginal} coupling. 
 Assuming a constant $\Phi$ in (1) we may further integrate $\vec{A}$
 to find the energy per unit volume to be
 \begin{eqnarray}
 S[\Phi] - S[0] = (\frac{\mu^2}{\Lambda^2} + \frac{N}{48})\Phi^2 +
 \frac{1}{2}(\frac{b}{\Lambda}- \frac{\pi^2 N^2}{48}) \Phi^4 + \\ \nonumber 
 + \frac{1}{6\pi^2} \ln( 1+ \frac{\pi^2 N}{4}\Phi^2)  + 
 \frac{\pi^4 N^3}{384}\Phi^6 \ln (1+ \frac{4}{\pi^2 N \Phi^2}), 
 \end{eqnarray}
 where we have rescaled $\Phi^2$, with the ultraviolet cuttof
 $\Lambda$ implicit in (1).
 Standard analysis of the Eq. (6) shows that there is a discontinuous
 transition for $ N > (4/\pi)\sqrt{2b/\Lambda}) $. Using the lowest order
 fixed point value $b/\Lambda = (2\pi^2 /5)(4-d) + O( (4-d)^2 )$
 in $d=3$ as a crude estimate of this bound,  we find the first-order
 transition for $N > 3.58 $, in rough agreement with
 \cite{dom}. Of course, this conclusion is to be trusted only
 for $N\gg 1$, when the first-order transition occurs at a large
 $\mu^2$, at which our neglect of fluctuations in $\Phi$ in arriving
 at the Eq. (6) becomes
 justified. For $N\ll 1$, on the other hand, the phase transition
 in the theory (1) should be continuous, with weakly modified XY exponents:
 $\nu=\nu_{xy}+O(N)$, $\eta=\eta_{xy}+O(N)$ \cite{dom}, \cite{lannert}.
 The situation is reminiscent of the Ginzburg-Landau superconductor 
 \cite{herb-tes}, with $N \gg 1$ analogous to the strongly type-I,
 and $N\ll 1$ to the extreme type-II case.   Of course, in dSC
 $N=2$ and fixed, so the issue is whether this falls into the type-I or into
 the type-II regime. In this context it 
 may be interesting to note that $T_c$ does appear to
 drop discontinuously to zero at the critical doping in LSCO \cite{fujita},
 and very steeply, if not even discontinuously, in YBCO \cite{doug}.

\section{Spin response in the superconductor}

\subsection{Derivation of the response function}

 Finally, we turn to spin dynamics deep inside the dSC, when $m/b \gg 1 $.
 For small momenta, $p \ll \Lambda_{Th}\approx 5 m$, we find that 
 $G^0 _{aa,\mu \nu} (p,m) = ( (\pi/24 m) +O(p^2) )
 (\delta_{\mu\nu}- \hat{p}_\mu \hat{p}_\nu )$. To keep the algebra simple,
 in the following we will approximate the gauge-field propagator with the 
 constant mass term, corresponding to the leading contribution in the
 above. The retention of the full form of $G^0 _{aa} (p,m) $
 should not qualitatively change the results, although it is more than
 likely that there will be some quantitative differences.

 Integrating out such a {\it massive} $\vec{a}$  in the superconducting
 phase leads to the effective $2+1$ dimensional Thirring model \cite{kondo}
 for spinons
 \begin{equation}
{\cal L} = \bar{\Psi}_i \gamma_\mu \partial_\mu  \Psi_i + 
\frac{\pi}{72 m} \bar{\Psi}_i \gamma_\mu \Psi_i
\bar{\Psi}_j \gamma_\mu \Psi_j,
\end{equation}
with the summation over repeated indices and the ultraviolet
cutoff $\Lambda_{Th}$ assumed. Using the
Hubbard-Stratonovich transformation \cite{negele}
we can rewrite this exactly as
\begin{eqnarray}
{\cal L} = \bar{\Psi}_i \gamma_\mu \partial_\mu  \Psi_i
+ \frac{18 m }{\pi} Tr [ M_{\mu} ^{ij} M_{\mu}^{ji}]  + 
Tr [ M_{\mu} ^{ij}\gamma_\mu \Psi_j \bar{\Psi}_i ]  
\end{eqnarray}
Making an {\it ansatz} $ M_{\mu}^{ij}(x) =
(M(x)/3) \delta_{ij} \gamma_\mu$ at the saddle point,
and then integrating out the fermions yields
\begin{equation}
S= N\int d^3 x [ \frac{24 m}{\pi} M^2 (x)
- Tr [ \ln ( \gamma_\mu \partial_\mu - M(x) ) ]   ]. 
\end{equation}
Expanding further in powers of $M(x)$ we finally write
\begin{equation}
\frac{S}{N}= \int \frac{d^3 q}{(2\pi)^3 }
( \frac{24 \pi m- \Lambda_{Th} }{\pi^2} + \frac{|q|}{8} )M^2 (q) + O(M ^4 ), 
\end{equation}
where $M^2 (q) = M(q) M(-q)$.

Using the definition of the Dirac fields in terms
of the electron creation and annihilation operators \cite{herbut} it
readily follows that 
\begin{equation}
\langle M(x) \rangle =
\frac{\pi}{12 Nm} \langle \hat{S}_z ( \vec{r},\tau) \rangle
\sum_{i=1}^2 \cos( 2 \vec{K}_i \cdot \vec{r}), 
\end{equation}
 where the vectors $\pm \vec{K}_i$, $i=1,2$ denote the positions of the four
nodes of the d-wave order parameter, as before.
$\langle M(q=0)\rangle $ is
therefore the static SDW OP \cite{herbut}, which, of course, was the
motivation behind our saddle-point ansatz. Similarly,
\begin{eqnarray}
\langle M^2 (\vec{q}, \omega) \rangle
- \frac{\pi}{48 Nm}=  \\ \nonumber 
( \frac{ \pi }{24N  m})^2 \sum_{i=1}^2
( \langle \hat{S}_z ( \vec{q} + 2\vec{K}_i, \omega)
\hat{S}_z ( -\vec{q} -2 \vec{K}_i, -\omega)\rangle +  \\ \nonumber
\langle \hat{S}_z ( \vec{q} - 2\vec{K}_i, \omega)
\hat{S}_z ( -\vec{q} + 2 \vec{K}_i, -\omega)\rangle).  
\end{eqnarray}
Both Eqs. (11) and (12) represent  exact equalities.

Since the dSC is rotationally invariant, the spin-spin correlation function
 is diagonal in indices $\alpha$, $\beta$,
 $\langle \hat{S}_\alpha ( \vec{k}, \omega)
\hat{S}_\beta ( -\vec{k}, -\omega) \rangle = \chi (\vec{k},\omega)
\delta_{\alpha \beta}$. We may therefore finally deduce that the imaginary
part of the spin response function $\chi(\vec{k},\omega)= \chi ' (\vec{k},\omega)
+ i \chi ''(\vec{k},\omega)$ in the phase fluctuating dSC is
\begin{equation}
\chi '' (2\vec{K}_i\pm \vec{q},\omega ) = (\frac{ 12 Nm}{\pi})^2
Im \langle M^2 (\vec{q}, \omega)\rangle. 
\end{equation}

Neglecting the quartic and higher order terms in Eq. (10) (which may be 
formally justified by assuming $N\gg1$),  and
analytically continuing to {\it real} frequencies $i\omega\rightarrow
\omega $, in the Gaussian approximation we finally obtain
\begin{equation}
\chi '' (2\vec{K}_i\pm \vec{q},\omega )  =
(\frac{8 m}{\pi})^2  \frac{ N \Theta(\omega^2 - q^2)
\sqrt{\omega^2 - q^2 } }
{ (64 m/\pi)^2 + \omega^2 - q^2}, 
\end{equation}
where we have written the cutoff $\Lambda_{Th}\approx 5 m$
as $\Lambda_{Th}= (16/\pi) m $, for convenience.  Eq. (14) 
is our central result. Its' most important feature is 
that the relativistic form of the quasiparticle dispersion near
the nodes forces $\chi''$ to vanish below a certain momentum-dependent
frequency.

First, let us check that deep in the superconducting phase we recover 
the result for the free quasiparticles. Taking $m\rightarrow \infty$
in the Eq. (14)
\begin{equation}
\lim_{m\rightarrow \infty} \chi '' (2\vec{K}_i\pm \vec{q},\omega )  =
\frac{ N}{64} \Theta(\omega^2 - q^2)
\sqrt{\omega^2 - q^2 }, 
\end{equation}
in agreement with \cite{rantner} and \cite{franzetal}
 for $N=2$. For $q=0$, for example,
$\chi '' \sim |\omega|$, and the spin response at $\pm 2\vec{K}_i $
is completely suppressed, as expected.

\begin{figure}
\centerline{\epsfxsize=7cm \epsfbox{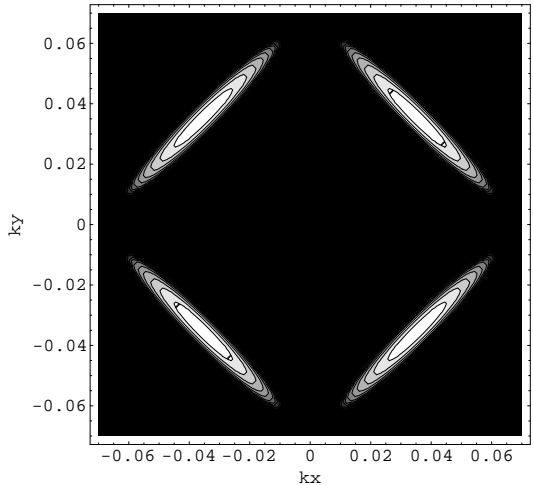}}
\vspace*{1em}

\noindent
\caption{ Contour plot of the $\chi ''$ at a low frequency
$\omega= 6 meV$, where the spin response peaks at four
diagonally incommensurate `mother' peaks. The origin
corresponds to $(\pi,\pi)$, and numbers are in units of $2\pi$. For other
parameters see the text. Brighter shade denotes a higher intensity. $\chi ''$
vanishes in the black regions.}
\label{f2}
\end{figure}

Next, we connect the uniform vortex susceptibility $m$ to 
the measurable superconducting $T_c$.
Assuming a continuous dSC-SDW transition,
$m^2 \propto (\mu^2 -\mu_c ^2)^{\gamma}$,
with $\gamma= \nu (2-\eta)$ characterizing the quantum critical point;
also, $T_c \propto (\mu^2 - \mu_c ^2 ) ^{z \nu}$. Since the dynamical
critical exponent in the theory (1) is $z=1$, 
$m\propto T_c ^{1- (\eta/2)}$. Sufficiently away from the critical
point we may neglect the anomalous dimension $\eta$ 
and find the result dictated by the (engineering)
dimensional analysis, $ m= c (\pi T_c/64)$, where $c$ is a
number. This would remain true even if the transition is weakly
first order. 

On the other hand, retaining the charge degrees of freedom one finds 
that the $T=0$ superfluid density is  $\rho_{sf} \approx 24m / \pi $,
in the underdoped regime \cite{igornew}. This
allows us to estimate the number $c$ by utilizing
the Uemura scaling \cite{uemura}, by which in our units
$\rho_{sf}\approx 2.1 T_c$ in
underdoped cuprates \cite{schneider1}. This number is known not to be
truly universal, but due to structural similarities between the cuprates
the variations from one material to another are reasonably
small \cite{schneider1}. The results that follow
will at any rate turn out to be quite insensitive to its exact value.
This finally yields $c\approx 5.6$, i. e. 
\begin{equation}
\frac{64 m}{\pi} \approx 5.6 T_c. 
\end{equation}
This should be understood as only a crude estimate; changing the cutoff 
$\Lambda_{Th}$, for example, will alter the numerical value of the 
constant $c$. This inherent limitation of the present
 calculation underlines the
need for a better treatment of the gauge field term in the
superconducting phase, that would improve upon our
constant mass approximation. This would, however, make our calculation
significantly more technical, and obscure somewhat the simple
physics we wish to address. Keeping these cautionary remarks in mind, we
proceed to observe the consequences of Eq. (14) in the following section.

\subsection{Evolution with frequency}

 We look for the maxima of the $\chi''(\vec{q},\omega)$
 in the momentum space at a fixed $\omega$. There are four  regimes
 to discern:

 1) For low frequencies $\omega$, the maximum values of $\chi'' $ are
  located at four "diagonally" {\it incommensurate}
 wave vectors $\pm 2\vec{K}_{1,2}$ (i. e. at $\vec{q}=0$).
 In Fig. 2 we plot such four well separated peaks at $2 \vec{K}_1$,
 $-2\vec{K}_1+ (2\pi,2\pi)$, $2\vec{K}_2+ (2\pi,0)$, and
 $-2\vec{K}_2 + (0,2\pi)$, where we have used the periodicity provided
 by the underlying lattice to shift all the peaks into the
 vicinity of the commensurate wave vector
 $\vec{Q}=(\pi,\pi)$. For illustration, we assumed $T_c = 67 K$ \cite{arai},
 $v_F / v_\Delta = 7$ \cite{krishana}, $v_F = 1.2 eV A$ \cite{valla},
 and estimated $2 \vec{K_1}= 0.465 (2 \pi,2 \pi) $ from
 \cite{dai}. These `mother' peaks
 are narrow and of low intensity, since both their width and the
 magnitude are controlled by the frequency.

\begin{figure}
\centerline{\epsfxsize=7cm \epsfbox{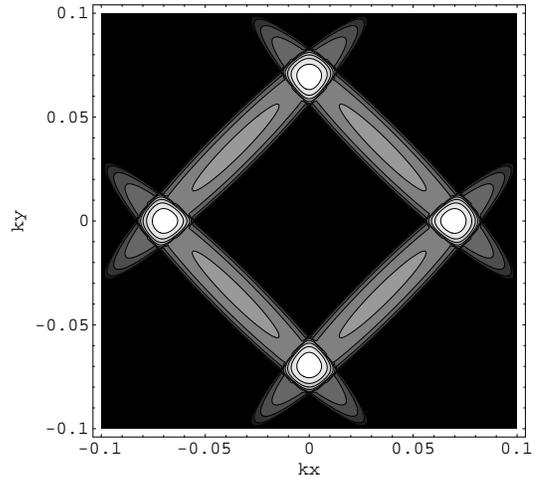}}
\vspace*{1em}

\noindent
\caption{The same as Fig. 2, but at a larger frequency, $\omega= 15 meV$. Note
that the maximum of the spin response is now found at the `parallel'
incommensurate wave vectors, because of the overlap of the mother peaks.}
\label{f3}
\end{figure}

 2) As the frequency
 increases both the the intensity and the width of the mother peaks
 in the momentum space grow. Since $v_F \gg v_\Delta$ they first start to
 overlap at four `parallel' positions (Fig. 3).  This way the superposition
  of the mother peaks leads to appearance of the maximum response at
  four incommensurate `parallel' wave vectors. We emphasize that this
  type of incommensuration has nothing to do here with any sort of 
  one dimensional (`stripe') ordering, as often assumed in literature. 

3) With a further increase of frequency the four initial peaks
 start overlapping at $\vec{Q}=(\pi,\pi)$,
 and for a while the `commensurate' response
 dominates (Fig. 4). With the above parameters we find the response at
$(\pi,\pi)$ starting developing at $\sim 60 meV$ (see Fig. 6).
The  energy of the `resonance' at which the response at
$\vec{Q}=(\pi,\pi)$ is
maximal is $\omega_{res}\approx 68 meV $. The overlap of all four mother
 peaks makes the response large in the whole interior of the square in Fig. 2. 

4) Finally, at the largest frequencies the maximum in Eq. (14) shifts to a
 $|\vec{q}| \neq 0$, which implies a weak redistribution of the
 commensurate peak to four `parallel' incommensurate positions again
 (Fig. 5). However, whereas the position of the `parallel'
 incommensurate peaks at Fig. 3 was
 independent of frequency at low frequencies, now it {\it increases} with
 frequency.

\begin{figure}
\centerline{\epsfxsize=7cm \epsfbox{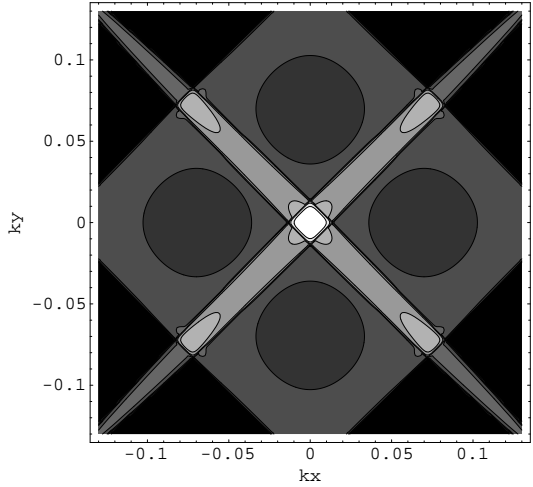}}
\vspace*{1em}

\noindent
\caption{Spin response at  $\omega= 72 meV$, near the resonance
energy of $\omega_{res}= 68 meV$. The overlap of all four
mother peaks from Fig. 2 has now produced the largest response at
the origin, i. e. at the
commensurate wave vector $\vec{Q}=(\pi,\pi)$.}
\label{f4}
\end{figure}

\subsection{Comparison with experiment}

 Let us review some of the features of neutron scattering
 experiments on underdoped cuprates pertaining to our discussion.
 We will concentrate on the underdoped YBCO at low temperatures,
 where the most detailed study of spin response as a function of
 frequency and wave vector exists \cite{arai}.

 1) At low frequencies ($\omega < 40 meV $) the maximum response is at
 four `parallel' incommensurate wave vectors \cite{mook}, \cite{arai}.
 The location of these peaks is very weakly dependent, or almost independent,
 on frequency \cite{arai}.

2) At a frequency $\omega\sim 40 meV$ the maximum shifts to the commensurate
position at $(\pi,\pi)$, and peaks there at a slightly larger
energy.  This is the so-called `resonance'.

3) At a higher frequency $\omega\sim 50 meV$ the resonance becomes flat,
and two weak `parallel' incommensurate peaks can be discerned again. This
time, however, their incommensuration {\it increases} with frequency.

4) The energy of the resonance decreases with the superconducting
 $T_c$ \cite{dai}, \cite{fong}.

5) At $\vec{Q}=(\pi,\pi)$ there is a `spin gap':
the commensurate response is strongly
suppressed below a certain energy $\omega_{sg}$.
At high frequencies the response also
continuously decreases, leaving a maximum at the `energy of the
resonance'.

6) Overall spin response increases with frequency at low frequencies,
peaks around the energy of the resonance, and then decreases with further
increase in frequency.

We see that the features 1)-3) are at least qualitatively
reproduced by our  Eq. (14),
once the lattice periodicity and the velocity anisotropy are taken
into account. Also, if we assume that $v_F$ and $v_\Delta$ are
only weakly doping dependent, the only energy scale left is
the gauge field mass $m$, which is connected to the doping dependent
superconducting $T_c$. The observed scaling of resonance energy with $T_c$
in underdoped samples therefore automatically follows from our
theory. Note that this feature otherwise  may appear quite
counterintuitive: if one believes that in underdoped cuprates
$T_c$ is the temperature where the phase coherence sets in, while the 
higher and increasing pseudogap temperature $T^*$ is related to the
amplitude of the OP, the observed connection between the spin response and
the superconducting phase coherence would seem rather mysterious.

The resonance at $\vec{Q}=(\pi,\pi)$,
 in our notation corresponds to some finite $\vec{q}_0=\vec{Q}-2\vec{K}_1$.
 This implies that as a function of frequency, at $T=0$ $\chi ''$ vanishes 
 below the spin gap $\omega_{sg} \approx v_F q_0 \approx 60 meV$,
 which is in our theory
 essentially determined by the incommensuration of the mother peaks and the
 magnitude of $v_F$ (Fig. 6). While our estimate appears to be too 
 large, our result naturally accounts for
 the existence of the spin gap. Also, the maximum response in Eq. (14) 
 is at the energy $\omega_{res} = \sqrt{\omega_{sg}^2 +(c T_c)^2} $ ,
 which {\it decreases} with decreasing $T_c$, as observed
 \cite{dai}, \cite{fong}. The experimental data of Dai et al. \cite{dai}
 are well reproduced by this simple formula with a just slightly smaller
 constant $c$ than in Eq. (16), and $\omega_{sg}=22 meV$ (Fig. 7). 
 As $T_c\rightarrow 0$, therefore, the commensurate
 peak energy $\omega_{res} \rightarrow \omega_{sg} \neq 0$.
We thus predict that the spin gap, and the
 $T_c \rightarrow 0$ extrapolated energy of the resonance  are essentially
 the same.  Remarkably, the spin gap  that can be estimated from a different
 set of data on YBCO (Fig. 17 in \cite{fong} )
 is quite similar, $\omega_{sg} \approx 20 meV$, consistent with our
 interpretation. At large energies, $\chi ''$ should behave as $\sim 1/\omega$
 (see Fig. 6), which also appears to be in a 
 qualitative, but not quite a quantitative, accord with the data.

\begin{figure}
\centerline{\epsfxsize=7cm \epsfbox{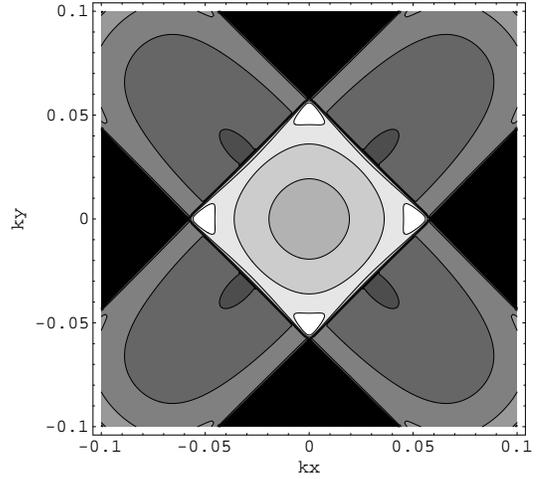}}
\vspace*{1em}

\noindent
\caption{Spin response at a high frequency $\omega= 110 meV$. Note the
weak redistribution of the intensity from the `resonance' in the center
to the `parallel' incommensurate peaks, which now will also
acquire an upward dispersion. }
\label{f5}
\end{figure}

 Finally, we predict a new feature of the spin response at the lowest
 frequencies: four weak and narrow `diagonally' incommensurate
 mother peaks, the  energy of which vanishes with
 $T_c$. Their low intensity and sharpness make them rather elusive,
 and indeed they have yet to be observed. Nevertheless,
 the fact that the evolution of the spin response
 with frequency as observed in underdoped YBCO can be straightforwardly
 understood in terms of these makes one optimistic about their
 existence. We find that the mother peaks should resolve as in Fig. 2
 at a frequency $\sim \omega_{res} /10$. Our optimism is further supported by
  the fine details of the measured
 spin response, which appear so far to be in accord with our picture.
 For example, Eq. (14)
 predicts that the response function at the `parallel' peaks must always be
 less than twice larger than that at $\pm 2\vec{K}_{1,2}$ at the same
 frequency. Closer inspection of the data in \cite{dai} or \cite{arai},
 for example, conforms to this expectation.

\begin{figure}
\centerline{\epsfxsize=7cm \epsfbox{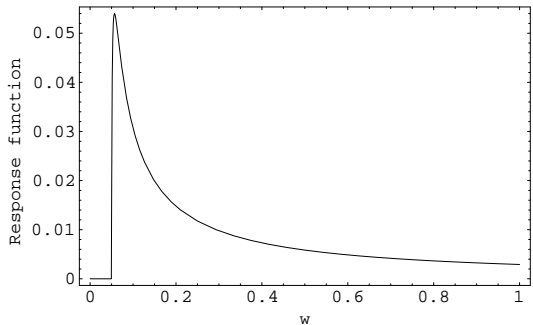}}
\vspace*{1em}

\noindent
\caption{ Spin response function $\chi''$ (in arbitrary units)
at the commensurate wave vector $\vec{Q}=(\pi,\pi)$ as a function of energy
(in units of $v_F= 1.2 eVA$). Note
that the spin response vanishes below certain frequency (`spin gap'),
sharply peaks just above it, and
behaves like $\sim 1/\omega$ at large frequencies.}
\label{f5}
\end{figure}

\section{Conclusion and discussion}

We have proposed an effective theory for the low-energy spin
excitations in the  strongly fluctuating d-wave superconductor.
The theory extends the $QED_3$ of the pseudogap
phase into the superconducting phase. It suggests that there is a
single dSC-SDW transition, which may be fluctuation
induced first order. We derived approximately the spin response function of
the fluctuating d-wave superconductor, and used it to qualitatively
explain the neutron scattering experiments in underdoped   YBCO.

Besides standing in clear opposition to most of the mean-field theories
of the high temperature superconductivity \cite{tremblay}, \cite{brinckmann},
our conclusion that there is no coexistence region between the dSC and the SDW
is also in contrast to the result of Pereg-Barnea and Franz \cite{pereg}, who
studied a closely related theory to (1). In their model 
spinons are coupled to the massive gauge field in the
superconducting phase, so that after integration over the gauge field
the model 
would become similar to the Thirring model in our Eq. (7). This leads to
the conclusion 
that the chiral symmetry for fermions  becomes broken at a {\it finite} 
mass of the gauge field $m$, which implies coexistence of the
SDW order and the phase coherence. One can readily understand this
result by considering the dimensionless coupling constant in the Thirring model
$g= \Lambda_{Th} /m$. Indeed, if the cutoff $\Lambda_{Th}$ is kept constant,  
$g\rightarrow g_c \sim 1$ at a finite $m$, and the model exhibits its 
well-known chiral instability. This result, however, in the more
complete theory (1) is avoided by remembering that in the Thirring model,
or in an equivalent description of \cite{pereg},
the cuttof needs to be $\Lambda_{Th} \sim m$ as well,
because for momenta $p \gg m$ the inverse gauge field propagator
becomes a linear function of $p$ (Fig. 1). With this modification, as
$m\rightarrow 0$, $g\rightarrow constant$, with the $constant$ 
 apparently too small for the chiral instability to occur {\it before} 
the critical point is reached.

 The result that superconducting and antiferromagnetic orders are not
comfortably 
coexisting in underdoped cuprates is also in agreement with
recent  experiments
on heat transport and c-axis penetration depth in clean underdoped YBCO
\cite{taillefer}, \cite{broun}. Both
measurements find that d-wave nodes survive in the underdoped regime,
without the quasiparticles becoming gaped. If there would be a
sufficiently strong 
coexisting antiferromagnetic ($\vec{Q}=(\pi,\pi)$) or SDW ($\vec{Q}
\neq (\pi,\pi)$) order, on the other hand, quasiparticles would acquire
precisely such a gap at the nodes. It should be said, however, that
the coexistence of SDW and SC orders is not in principle impossible,
and can be induced by a strong favourable interaction term, for example, 
omitted in (1). Our point is only that vortex condensation as the 
dynamical mechanism behind the appearance of the  SDW order in the 
theory (1) does not lead to such coexistence. 

\begin{figure}
\centerline{\epsfxsize=7cm \epsfbox{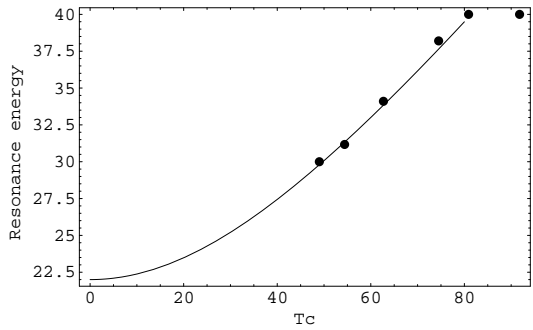}}
\vspace*{1em}

\noindent
\caption{The energy of the resonance (in meV)
vs. the superconducting $T_c$ (in K). Data are from
the Fig. 29 in second ref. 14, and the
fit is to $\omega_{res} = \sqrt{ (c T_c)^2 + \omega_{sg} ^2}$
(see the text), with $c = 4.75$, and $\omega_{sg}= 22 meV$.}
\label{f5}
\end{figure}

 Our picture of spin response in underdoped cuprates is similar in
 spirit to that of Rantner and Wen \cite{rantner}. There are important
 conceptual and formal differences, however.  Rantner and Wen considered  the
 pseudogap regime, {\it above} the superconducting $T_c$, and proposed
 that spinons may there effectively be described as coupled to a
 massless gauge field. Neglecting the chiral instability such a theory
 is almost
 universally believed to have, they proposed the algebraic spin liquid
 as the pseudogap phase, in which
 in the large-N approximation the spin response is  in fact 
 {\it suppressed} at low frequencies. Only if the number of fermion
 components N is taken sufficiently small (smaller in fact than the $N_c$ 
 below which the chiral instability occurs in the
 same calculation) does one recover the enhancement
 of the spin response.  To account phenomenologically
 for the phase coherence in the
 superconducting state Rantner and Wen added a mass to their
 gauge field, which then further suppresses the spin response at  energies
 below the mass. We see that a qualitatively similar mechanism
 operates in our theory, but at $T=0$: the gauge field coupled to spinons is
 massless outside the SC, and massive inside.
 The form of our spin response function
 $\chi ''$, however, is quite different, which, for example,
 leads to reappearance of the incommensurate peaks at high energies
 (Fig. 5), among other things. Most importantly, the mass
 of the gauge field $m$ in the superconducting phase that was postulated
 in \cite{rantner} on phenomenological grounds here becomes a {\it result}
  within the effective theory (1).

  With the parameters we used the energy of the resonance
  $\omega_{res} \sim 70 meV$, not far from the generic $\sim 40 meV$
see in YBCO, and more recently in $Bi_2 Sr_2 CaCu_2 O_{8+\delta}$ (BSCCO)
\cite{fong1}, \cite{he}. Given the crudeness of our
approximation this is an encouraging agreement. However, there is a
considerable uncertainty in the parameters we used which may easily
make this agreement fortuitous. For example, assuming
$v_F= 1.8 eV A$ \cite{johnson} would push the resonance energy in our
calculation to $\sim 100 meV $. Also, we do not find the $\delta$-function
response at the commensurate wave vector, but only a $1/\omega$ decay
on the high energy side (Fig. 6). This is contrast to the
mean-field calculation within the $t-J$ model of Brinckmann and Lee
\cite{brinckmann}, who found that the next-nearest-neighbor
hopping may lead to a $\delta$-function term in $\chi''$. Our result,
however, seems to agree with the measurements of Fong et al. \cite{fong}
on underdoped YBCO, where the width of the resonance
of $\sim 10 meV$ was observed (see Fig. 7 in \cite{fong}).
It is interesting that Brinckmann and Lee, just like us, also tend to 
overestimate the energy of the resonance, and found it necessary to
reduce the coupling $J$ by hand by a factor of three in their
calculation to match the experiment.

We find the commensurate response to dominate only within a certain window
of energies around the resonant energy, while both at lower
($\omega<\omega_{sg}$) and higher ($\omega > 1.5 \omega_{sg} $) 
energies the peaks are located at the parallel incommensurate
wave vectors. In our calculation the resonance 
exists for $60 meV < \omega < 90 meV$, with the parameters we used.
For a  lower $T_c$ the window of the dominant response at $\vec{Q}=(\pi,\pi)$
shrinks. This may provide an explanation why the resonance was never found
in LSCO, but has  been  seen in both YBCO and BSCCO
which both have a higher $T_c$, without other parameters being too 
different. Alternatively, the physics of underdoped LSCO may be
qualitatively different, as some measurements would suggest \cite{fujita}. 

We end by repeating the central prediction of this work: the appearance of
four weak and narrow `diagonal' incommensurate peaks, the energy of which
should extrapolate to zero with vanishing superconducting $T_c$. It is
the interplay of these fundamental peaks that leads to 
evolution of spin response with frequency in our calculation in qualitative
agreement with experiment. While some other theories
 \cite{brinckmann}, \cite{rantner} would also lead to such `diagonal'
incommensuration at low frequencies, we also predict that the energy
of these peaks should vanish with $T_c$, since they represent
the intrinsic soft mode of the phase fluctuating d-wave superconductor.
Detection of this feature in future neutron scattering experiments would
provide strong support in favor of the theory of the underdoped cuprates
advocated here.

\section{Acknowledgement}

 This work was supported by NSERC of Canada and the Research Corporation.
IFH also thanks Aspen Center for Physics for its hospitality, and B.
Seradjeh for useful discussions.

\end{document}